\documentclass[10pt]{iopart} 

\usepackage[alsoload=synchem]{siunitx}
\usepackage{graphicx}

\usepackage[colorlinks=true, allcolors=blue]{hyperref}

\begin{document}

\title{Growth and Characterization of Homoepitaxial $\beta$-Ga$_2$O$_3$ Layers}

\author{M. Brooks Tellekamp, Karen N. Heinselman, Steve Harvey, Imran Khan, Andriy Zakutayev}
\address{National Renewable Energy Laboratory, 15013 Denver West Blvd., Golden, Colorado, USA, 80401}
\ead{brooks.tellekamp@nrel.gov, andriy.zakutayev@nrel.gov}

\pagestyle{plain} 
\setcounter{page}{1} 

\begin{abstract}
$\beta$-Ga$_2$O$_3$ is a next-generation ultra wide bandgap semiconductor (E$_g$ = \SIrange{4.8}{4.9}{\eV}) that can be homoepitaxially grown on commercial substrates, enabling next-generation power electronic devices among other important applications. Analyzing the quality of deposited homoepitaxial layers used in such devices is challenging, in part due to the large probing depth in traditional x-ray diffraction (XRD) and also due to the surface-sensitive nature of atomic force microscopy (AFM). Here, a combination of evanescent grazing-incidence skew asymmetric XRD and AFM are investigated as an approach to effectively characterize the quality of homoepitaxial $\beta$-Ga$_2$O$_3$ layers grown by molecular beam epitaxy at a variety of Ga/O flux ratios. Accounting for both structure and morphology, optimal films are achieved at a Ga/O ratio of $\sim$1.15, a conclusion that would not be possible to achieve by either XRD or AFM methods alone. Finally, fabricated Schottky barrier diodes with thicker homoepitaxial layers are characterized by \textit{J-V} and \textit{C-V} measurements, revealing an unintentional doping density of \num[scientific-notation = true]{4.3e16} \si{\per\centi\meter\cubed} - \num[scientific-notation = true]{2e17} \si{\per\centi\meter\cubed} in the epilayer. These results demonstrate the importance of complementary measurement methods for improving the quality of the $\beta$-Ga$_2$O$_3$ homoepitaxial layers used in power electronic and other devices.
\end{abstract}


\maketitle
\ioptwocol

\section{Introduction}
$\beta$-Ga$_2$O$_3$ is a promising ultra wide band gap semiconductor\cite{higashiwaki_mbe_2019, tsao_ultrawide-bandgap_2018}, featuring 4.8 - 4.9 eV bandgap\cite{peelaers_brillouin_2015, orita_deep-ultraviolet_2000}, controllable n-type doping with group IV elements (Si, Ge, Sn)\cite{ahmadi_ge_2017, krishnamoorthy_modulation-doped_2017, sasaki_mbe_2013}, and a large-size low-cost\cite{reese_how_2019} bulk single crystal substrate wafers\cite{galazka_czochralski_2010, aida_growth_2008}. Because of these properties, $\beta$-Ga$_2$O$_3$ and related alloys\cite{wenckstern_group-iii_2017} are studied for solar-blind ultraviolet (UV) photodetectors\cite{qin_review_2019}, high-power and radio-frequency electronic devices \cite{chabak_lateral_2019, higashiwaki_state---art_2017}, high-temperature gas sensors\cite{afzal_-ga2o3_2019} and other potential applications\cite{pearton_review_2018}. In the field of $\beta$-Ga$_2$O$_3$ power electronic applications, research activities have rapidly advanced over the past decade leading to improved device performance\cite{ahmadi_demonstration_2017, moser_ge-doped_2017, liu_review_2019}. For example, vertical and trench Schottky barrier diodes have been achieved with a Baliga's figure of merit (BFOM) approaching 1 \si{\giga\watt\per\centi\meter\squared}\cite{li_field-plated_2020, allen_vertical_2019}, and vertical field effect transistors with breakdown voltage $>$1 \si{\kilo\volt} and on-state resistance of \SIrange{15}{30}{\milli\ohm\centi\meter\squared} have been demonstrated\cite{wong_current_2019, hu_enhancement-mode_2018}. However, there are remaining and well-founded concerns regarding the low thermal conductivity\cite{slomski_anisotropic_2017, cheng_thermal_2019} and lack of p-type dopability of $\beta$-Ga$_2$O$_3$\cite{kyrtsos_feasibility_2018, peelaers_deep_2019}, which would have to be addressed for future power electronic device applications.

Despite the advances in $\beta$-Ga$_2$O$_3$ device fabrication technology, routine characterization of homoepitaxial layer quality is complicated due to difficulty in discriminating between the substrate and the epilayer. This has been especially challenging  when studying thin epitaxial layers in complex multi-layer device structures. Conventional x-ray diffraction (XRD) $2\theta-\omega$ scans are routinely performed, but their analysis is limited to determining the layer thickness from Pendell{\"o}sung interference fringes\cite{okumura_systematic_2014}. XRD rocking curves are commonly measured, but the full width at half maximum (FWHM) is dominated by the signal from the substrate\cite{sasaki_growth_2014}. Other commonly used method are sensitive only to the surface quality of the homoepitaxial layer, like reflection high energy electron diffraction (RHEED)\cite{sasaki_mbe_2013, okumura_systematic_2014} or atomic force microscopy (AFM)\cite{sasaki_mbe_2013, oshima_surface_2008}. In-plane grazing incidence X-ray diffraction (GIXD) is a useful method to evaluate homoepitaxial layer quality\cite{lee_homoepitaxial_2016, peelaers_deep_2019}, however this technique requires a specialized diffractometer with an in-plane arm that is not as widely available to the scientific community as a standard 4-circle x-ray diffractometer. X-ray topography is another method capable of characterizing homoepitaxial layers, however it requires access to a synchotron radiation facility\cite{doi:10.1063/1.5051633}.

\begin{figure}
    \centering
    \includegraphics[width=8cm]{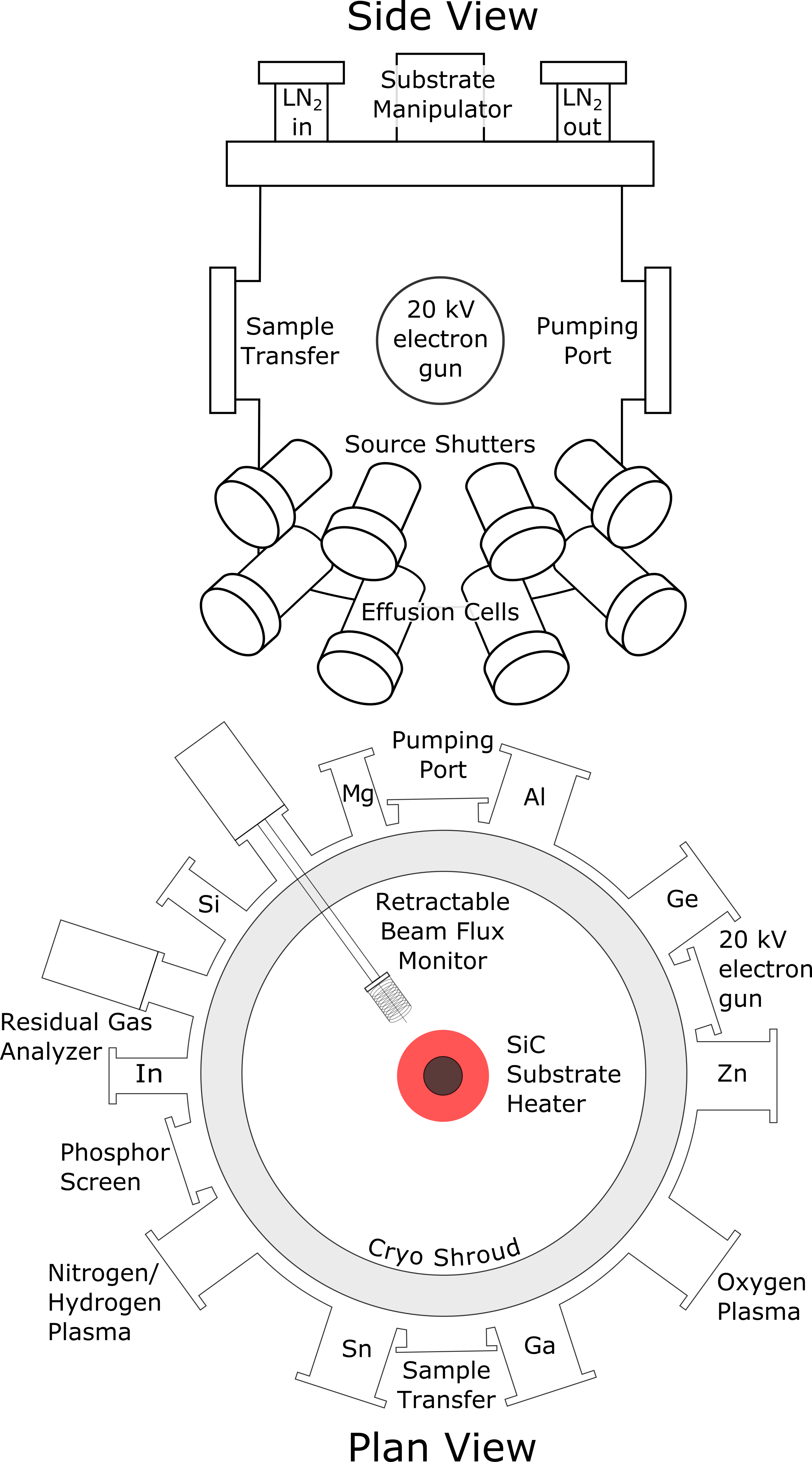}
    \caption{Plan view and side view schematics of the oxide/nitride MBE chamber used to grow $\beta$-Ga$_2$O$_3$ homoepitaxial layers. LN$_2$ - liquid nitrogen.}
    \label{fig:chamber}
\end{figure}

Here, we explore skew asymmetric X-ray diffraction (XRD) at grazing incidence angles as a method to quantify $\beta$-Ga$_2$O$_3$ homoepitaxial layer crystal quality. This method is more broadly accessible to the scientific community than GIXD through the use of a standard 4-circle diffractometer. The skew asymmetric XRD results are correlated with changes in surface morphology as measured by atomic force microscopy (AFM) across a wide range of Molecular Beam Epitaxy (MBE) growth conditions to determine optimal growth parameters. Secondary Ion Mass Spectrometry (SIMS) is used to verify the chemical purity of the epilayers and their thickness, obtained from fitting of interference fringes in XRD $2\theta-\omega$ scans to determine a growth rate curve. Finally, thicker homoepitaxial layers processed into vertical Schottky diodes are  characterized by current - voltage measurements to verify the  blocking characteristics of the homoepitaxial layer, and by capacitance - voltage measurements to determine the effective background electron concentration of the epilayer. Together, these results present a broader view of the role that growth stoichiometry plays in homoepitaxial layer quality.

\section{Experimental Methods}

$\beta$-Ga$_2$O$_3$ was grown homoepitaxially on (010) and ($\bar{2}$01) oriented 10 \si{\milli\meter} x 15 \si{\milli\meter} wafers (Tamura Corporation) using a Riber Compact 21T molecular beam epitaxy (MBE) deposition system which is schematically depicted in Figure~\ref{fig:chamber}. Substrates were cleaned by a sequential solvent rinse of acetone/methanol/2-propanol followed by a deionized (DI) water rinse. The substrates were then indium bonded to Si carrier wafers to absorb IR radiation from the substrate heater. 7N Ga was sourced from a SUMO style effusion cell, and high purity dry oxygen (SAES purifier, $<$ 1 part per billion H$_2$O by volume) was supplied through a 13.56 \si{\mega\hertz} radio-frequency oxygen plasma source operating at 300 \si{\watt} with flows of 2.0 SCCM - 3.0 SCCM. Substrates were outgassed in an introductory vacuum chamber at 150 \si{\celsius} for 90 minutes before transferring to an intermediate holding chamber. Before growth, Ga fluxes were measured using a Bayard-Alpert style retractable ionization gauge. All substrates were annealed at 800 \si{\celsius}, while exposed to an oxygen plasma operated at 300 W and 2.0 SCCM, for 30 minutes before growth. All films were grown at a substrate temperature of 700 \si{\celsius} as measured by a thermocouple on the opposite side of the carrier wafer, resulting in film thickness in the 100 nm range. Detailed film thicknesses can be found in the Supporting Information, Table 1. Films were analyzed \textit{in situ} by reflection high energy diffraction (RHEED).

XRD was performed using a Rigaku Smartlab diffractometer using Cu-K$\alpha$ radiation monochromated by a 4-bounce Ge-(220) crystal, and a Panalytical MRD Pro using Cu-K-$\alpha$ radiation with a hybrid monochromator (G{\"o}bel mirror plus a 4-bounce Ge-(400) crystal) on the incident beam. A Veeco/Bruker D3100, with a Nanoscope 5 controller unit, was used to collect AFM images.  Measurements were conducted tapping mode with a 20 \si{\nano\meter} at a scan rate of 0.35 \si{\hertz}, and the samples were secured using a vacuum stage. Post processing of images included flattening with a 3rd order function and a denoise algorithm.

An ION-TOF TOF-SIMS V Time of Flight SIMS (TOF-SIMS) spectrometer was utilized for depth profiling. Analysis was completed utilizing a 3-lens 30 \si{\kilo\volt} BiMn primary ion gun. Depth profiles were completed with the 30 \si{\kilo\eV} Bi$^+$ primary ion beam, (0.8 \si{\pico\ampere} pulsed beam current), a 50 \si{\micro\meter} $\times$ 50 \si{\micro\meter} area was analyzed with a 256:256 pixel primary beam raster. Sputter depth profiling for positive ion yield was accomplished with 3 \si{\kilo\eV} oxygen ion sputter beam (30 \si{\nano\ampere} sputter current) with a raster of \SIrange{150}{150}{\micro\meter}. Sputter depth profiling for negative ion yield was accomplished with 3 \si{\kilo\eV} cesium ion sputter beam (34 \si{\nano\ampere} sputter current) with a raster of \SIrange{150}{150}{\micro\meter}. 

\begin{figure}
    \centering
    \includegraphics[width=8cm]{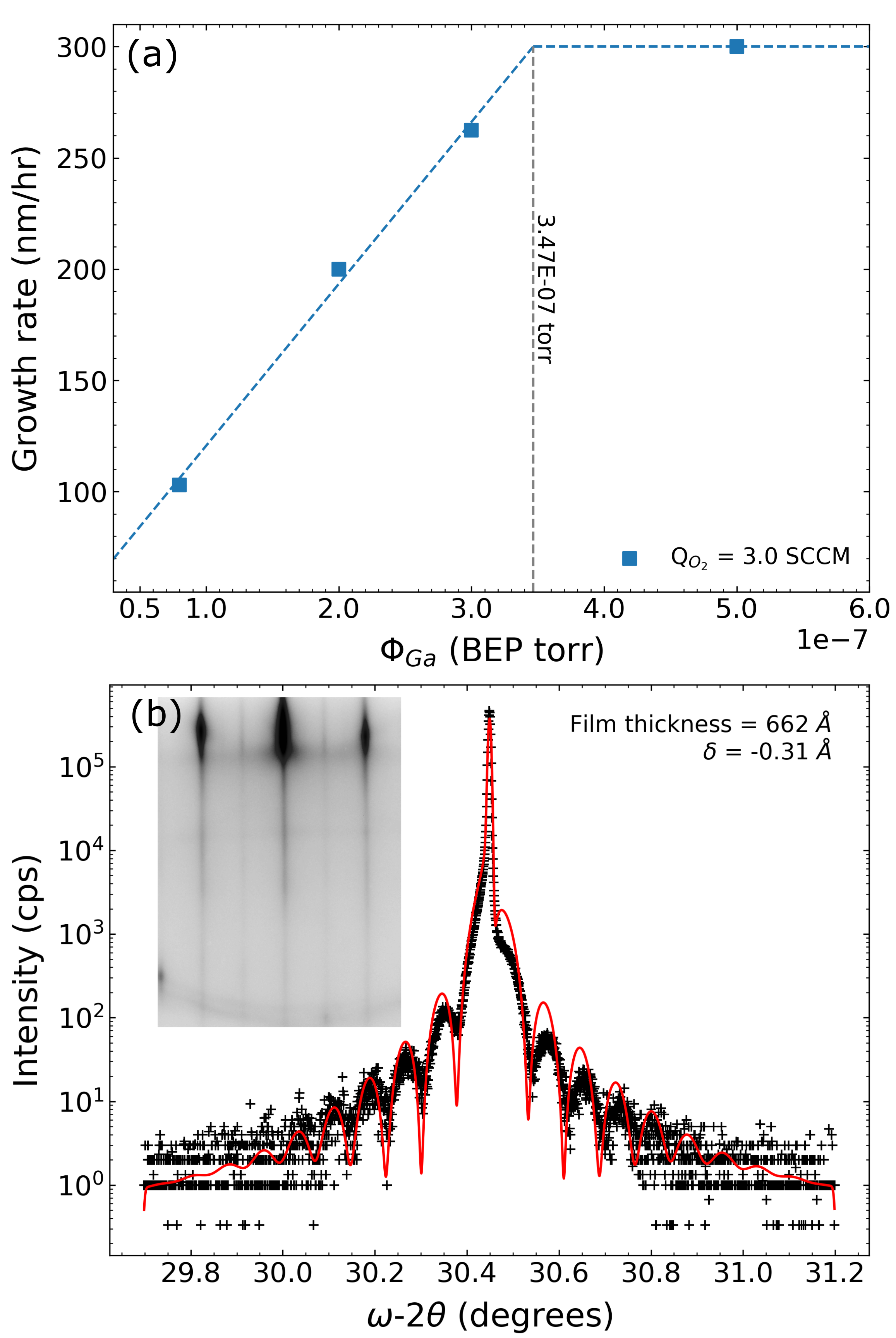}
    \caption{(a) Growth rate as a function of Ga flux measured in BEP torr. The growth rates are derived from XRD thickness fringes shown in (b), where an interface offset $\delta$ allows fringes to be observed. The oxygen flux corresponds to a growth rate of 300 nm/hr which is approximately equal to a Ga BEP of \num{3.47E-7} \si{\torr}. A typical RHEED image with the electon beam perpendicular to the $<$001$>$ direction is shown in the inset of (b).}
    \label{fig:xrd_stoich}
\end{figure}

Schottky contacts were deposited on thicker (~500 nm) MBE-grown $\beta$-Ga$_2$O$_3$ on Sn-doped ($\bar{2}01$) oriented substrates, and Ohmic contacts were deposited on the backside of the wafers to form vertical Schottky diodes. Post-growth, the wafers were cleaned via a rinse in 30\% hydrochloric acid (HCl) to remove the indium from the backside. Wafers were then cleaned with a quick succession of rinses in acetone, 2-propanol, and DI water, which was then dried with  N$_2$. Samples were then rinsed once again in 30 \% HCl, DI water, and dried with N$_2$ to prepare the backside surface for the ohmic contact. The backside ohmic contact, consisting of Ti(20 \si{\nano\meter})/Au(100 \si{\nano\meter}) was deposited by e-beam evaporation. Samples were then processed via rapid thermal annealing (RTA) in N$_2$ at 400 \si{\celsius} for 1 minute. A second clean of acetone, 2-propanol, and DI water, and dry N$_2$ was used prior to the e-beam evaporation of the Pt(50 \si{\nano\meter})/Ti(10 \si{\nano\meter})/Au(100 \si{\nano\meter}) circular Schottky contacts, which were defined by a shadow mask.

\begin{figure}
    \centering
    \includegraphics[width=7.5cm]{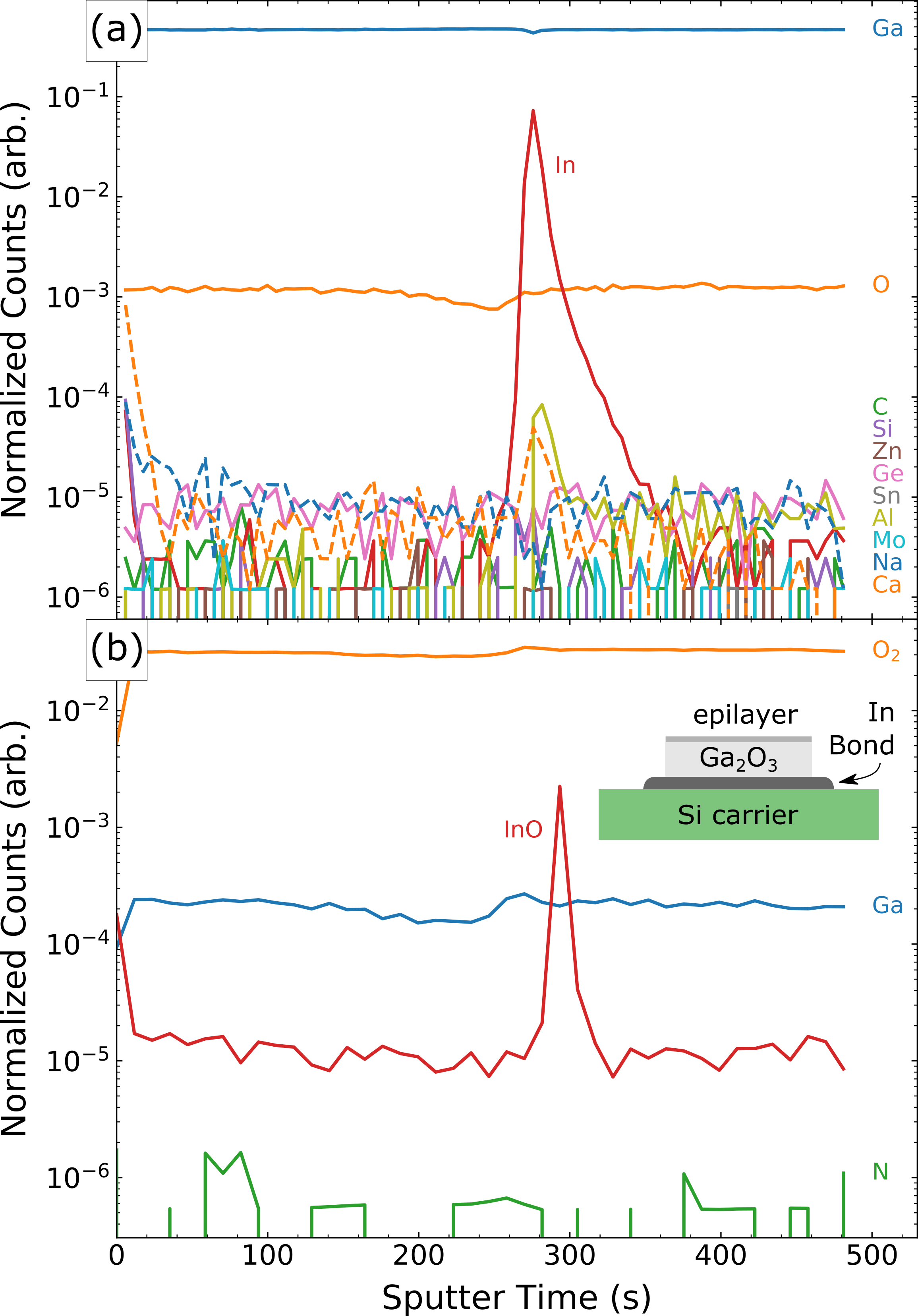}
    \caption{(a) Positive ion and (b) negative ion SIMS depth profiles of a $\beta$-Ga$_2$O$_3$ epilayer grown on an unintentionally doped substrate. Common impurities and chamber-specific impurities are not observed in the epilayer. A spike of In, Ca, and Al is observed at the homoepitaxial interface, likely due to In-bonding used to mount the substrate to a carrier wafer and due to residual Ca and Al impurities in SIMS chamber. The location of the In-bond is shown in the layer stack inset of (b).}
    \label{fig:sims}
\end{figure}

The Schottky diode characteristics were measured using a Keithley 4200A-SCS parameter analyzer. Current-Voltage (\textit{J-V}) data was taken in two steps, with the voltage scanned from \SIrange[tophrase={ -- $\pm$}]{0}{5}{\volt} and the current compliance set at 100 \si{\milli\ampere}. The on-off ratio, turn-on voltage, ideality factor, and on-state resistance were determined by fitting portions of the \textit{J-V} curves to diode equation. The Capacitance-Voltage (\textit{C-V}) data was taken at a frequency of 1 \si{\mega\hertz}, over a voltage range from \SIrange{-5}{5}{\volt}. The carrier concentration was determined from  \textit{C-V} using Mott-Schottky analysis at negative voltages where the capacitance was not limited by turn on of the diode. Analysis of the depletion width was limited to $sim$ 380 nm by the applied DC bias. To prepare the data points for differential analysis (d(1/C$^2$)/dV) the data was smoothed using a piecewise Savitsky-Golay filter of order 2, kernel 3 for 1 $<$ V $<$ 1.5 and order 3, kernel 11 for the remaining data.

\section{Results and Discussion}
\subsection{Film Growth}

To establish growth regimes and the stoichiometric growth rate, $\beta$-Ga$_2$O$_3$ films were initially grown at varying incident Ga fluxes. Using plasma conditions of 3.0 SCCM O$_2$ and 300 \si{\watt}, the stoichiometric oxygen flux was determined to be 300 nm/hr, corresponding to a Ga beam equivalent pressure (BEP) of \num{3.47E-7} \si{\torr}. This trend is shown in Figure~\ref{fig:xrd_stoich} (a), and is well aligned with the growth regimes previously reported for $\beta$-Ga$_2$O$_3$\cite{okumura_systematic_2014, ahmadi_demonstration_2017}. All film thicknesses were determined by analysis of Pendell{\"o}sung thickness fringes, as shown in Figure~\ref{fig:xrd_stoich} (b). While a perfect homoepitaxial film should not show these fringes, it has been well documented that a slight interface offset, $\delta$, is responsible for the appearance of these fringes\cite{okumura_systematic_2014, lebeau_stoichiometry_2009}. For the fit, instrument broadening is accounted for using a convolved gaussian, and fringe decay is modeled by surface roughness using a gaussian envelope. Films were also monitored \textit{in situ} by RHEED, and a representative pattern perpendicular to the $<$001$>$ azimuth is shown in the inset of Figure~\ref{fig:xrd_stoich} (b) indicating a smooth and high quality surface by the streaky nature of the pattern and the appearance of a 2$\times$ surface reconstruction, visible as a half order streak between the primary streaks.

\begin{figure*}
    \centering
    \includegraphics[width=15cm]{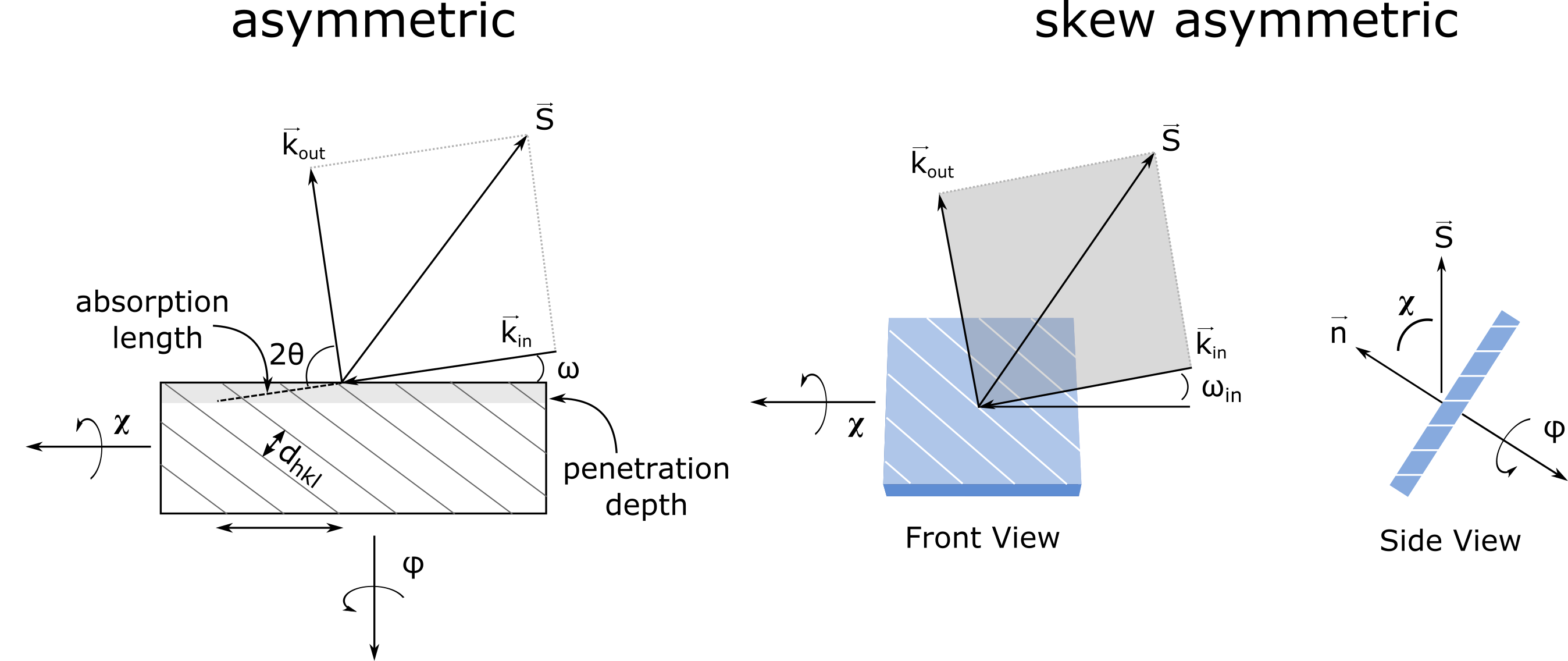}
    \caption{Asymmetric and skew asymmetric geometries used in XRD. Asymmetric geometry refers to any diffraction scan where $\omega$ = $\theta$ - offset. The penetration depth is then defined by the angle of the chosen diffraction planes. Skew geometry refers to a perpendicular sample rotation around the $\chi$ axis where the sample normal $\vec{n}$ is not within the scattering plane (shown in grey) defined by the scattering vector $\vec{S}$ and wavevectors $\vec{k_{in}}$ and $\vec{k_{out}}$. In skew asymmetric geometry, a grazing incidence angle ($\omega_{in}$) may be chosen for any arbitrary asymmetric diffraction condition to limit the penetration depth. Through coincident rotation of $\chi$ and $\phi$ angles, the lattice planes of interest can be brought in to alignment with the Bragg condition.}
    \label{fig:geometry}
\end{figure*}

 SIMS depth profiles were performed to investigate the origin of the interfacial offset resulting in the thickness fringes described above. Additionally, the MBE chamber used to deposit $\beta$-Ga$_2$O$_3$ films is also used to deposit other semiconductor materials, and SIMS can be used to analyze epilayers for these contaminants at impurity concentrations. Namely, the chamber contains Si, Al, Ge, Zn, In, Sn, O and N (see Figure~\ref{fig:chamber}). Mo blocks are also used to hold the Si carrier wafers in the chamber which could provide an additional source of contamination through the formation of volatile MoO$_3$\cite{stull_vapor_1947}. Figure~\ref{fig:sims}~(a) shows a SIMS depth profile analyzing positively charged ions highlighted above along with other common impurities including C and Na. Nitrogen was interrogated by a separate depth profile analyzing negative ions, shown in Figure~\ref{fig:sims}~(b). All bulk impurities, including Zn and O, were below the detection limit in the epilayer; however, contamination was observed at the homoepitaxial interface. In, Al, and Ca are observed to spike at the interface. The most likely source of In is the In-bond used to mount the substrates to carrier wafers, while the other signals are likely due to other samples measured in SIMS instrument prior to this one. A schematic of the bonding procedure is shown in the inset of Figure~\ref{fig:sims}~(b).

\subsection{Structural measurements}
XRD is the standard method for analyzing the structural uniformity of single crystals, however due to the large penetration depth of X-rays in solids compared to thin film thicknesses XRD is unsuitable for the characterization of homoepitaxial thin films--the diffracted intensity is dominated by the substrate. For $\beta$-Ga$_2$O$_3$, the vertical penetration depth of Cu-K$\alpha$ radiation is on the order of 10's of \si{\micro\meter}, while typical thin films are only 100's of \si{\nano\meter} thick, making traditional XRD analysis impossible. Alternative geometries can, however, be considered. Asymmetric geometry refers to an offset in the scattering vector $\vec{S}$ from the surface normal $\vec{n}$ resulting in different incident and exit angles, as shown in Figure~\ref{fig:geometry}. Skew geometry refers to a perpendicular tilt of the sample with respect to the incident beam direction, where $\vec{n}$ does not lie within the scattering plane defined by $\vec{k_{in}}$ and $\vec{k_{out}}$ (represented by the grey plane in Figure~\ref{fig:geometry}). In a skew asymmetric geometry, grazing incidence angles may be chosen to geometrically limit the penetration depth to smaller values, effectively attenuating the incident beam within the epilayer\cite{swiatek_x-ray_2017}. This is schematically depicted in Figure~\ref{fig:geometry}, where the incidence angle $\omega_{in}$ can be chosen by coincident rotations of the sample along the $\chi$ and $\phi$ axes at a fixed $\vec{S} = \vec{k}_{out} - \vec{k}_{in}$.

\begin{figure*}
    \centering
    \includegraphics[width=16cm]{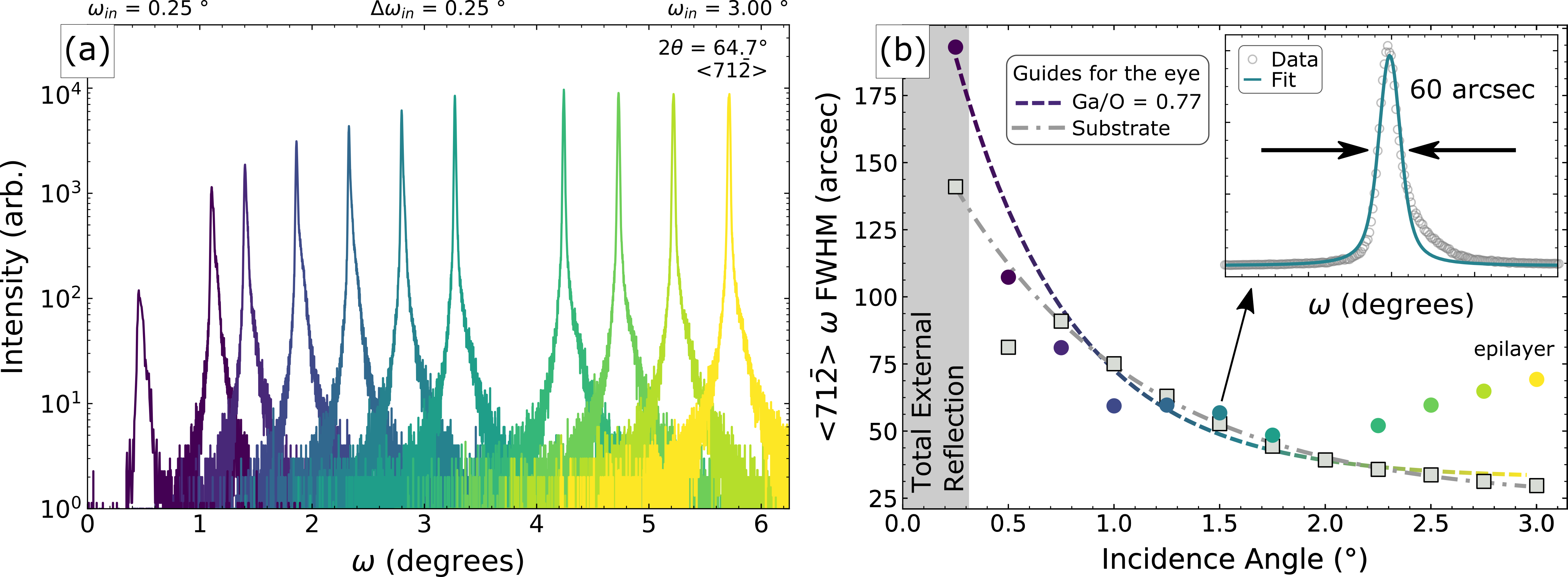}
    \caption{Skew asymmetric XRD of homoepitaxial $\beta$-Ga$_2$O$_3$. Pane (a) shows $\omega$ rocking curves as a function of $\omega_{in}$ for the 71$\bar{2}$ reflection. At low angles the penetration depth is severely limited. (b) shows the FWHM of each peak versus $\omega_{in}$, where the color of each data point matches to the corresponding raw data in (a). For increasing incidence angle the rocking curve FWHM decreases as signal from the substrate begins to dominate and coherence length broadening is decreased. By choosing a sufficiently grazing incidence angle, the penetration depth can be limited to the epilayer. Representative data from a bare substrate is also shown in (b) (gray squares), exhibiting broadening as the incidence angle is increased. Below the critical angle for total external reflection the deviation between substrate and an epilayer grown at a Ga/O ratio of 0.77 is evident. Guides to the eye are included. Slight misalignment as the measurement progresses to higher incidence angles are observed as a decrease in total intensity in (a) and a slight increase in FHWM (b). Guides to the eye assume convergence to the measured substrate value of $\sim$30 arcsec. An example fit with a Voigt profile is shown in the inset of (b). Fitting errors are smaller than the data point size.}
    \label{fig:xrd_diagram}
\end{figure*}

To study structural quality of the epilayers, the normally (010) oriented $\beta$-Ga$_2$O$_3$ films are analyzed by skew asymmetric XRD at grazing incidence angles along the (71$\bar{2}$) reflection, chosen for it's moderate inclination and strong scattering factor. Figure~\ref{fig:xrd_diagram} (a) shows $\omega$ rocking curves with incidence angles from \SIrange[tophrase={ -- }]{0.25}{3.0}{\degree} for a 185 \si{\nano\meter} thick epilayer grown at a Ga/O ratio of 0.77. The goniometer angles used to achieve these measurements are tabulated in the Supplemental Information, Table 2. At low incidence angles, the signal is dominated by the epilayer, but by increasing the incidence angle the penetration depth is rapidly increased, effectively probing the substrate. The resulting full width at half maximum (FWHM), extracted by fitting each rocking curve to a Voigt profile (Figure~\ref{fig:xrd_diagram} (b) inset), is plotted in Figure~\ref{fig:xrd_diagram} (b). The fit error is at most $\pm$ 3\si{\arcsec} and is smaller than the size of the plotted data points. This epilayer has a rocking curve FWHM of 193\si{\arcsec} at a 0.25\si{\degree} incidence angle. It should also be noted that measurements over a wide range of incidence angles, such as that shown in ~Figure~\ref{fig:xrd_diagram}, are automated due to long scan times and a requirement to re-optimize at each peak. The data point at 2 degrees incidence angle is omitted because the automation did not successfully optimize the peak and only measured noise. Similar but smaller misalignment issues at higher incidence angles led to a slight decrease in total intensity and a slight increase in FHWM in Figure~\ref{fig:xrd_diagram}.

\begin{figure*}
    \centering
    \includegraphics[width=16cm]{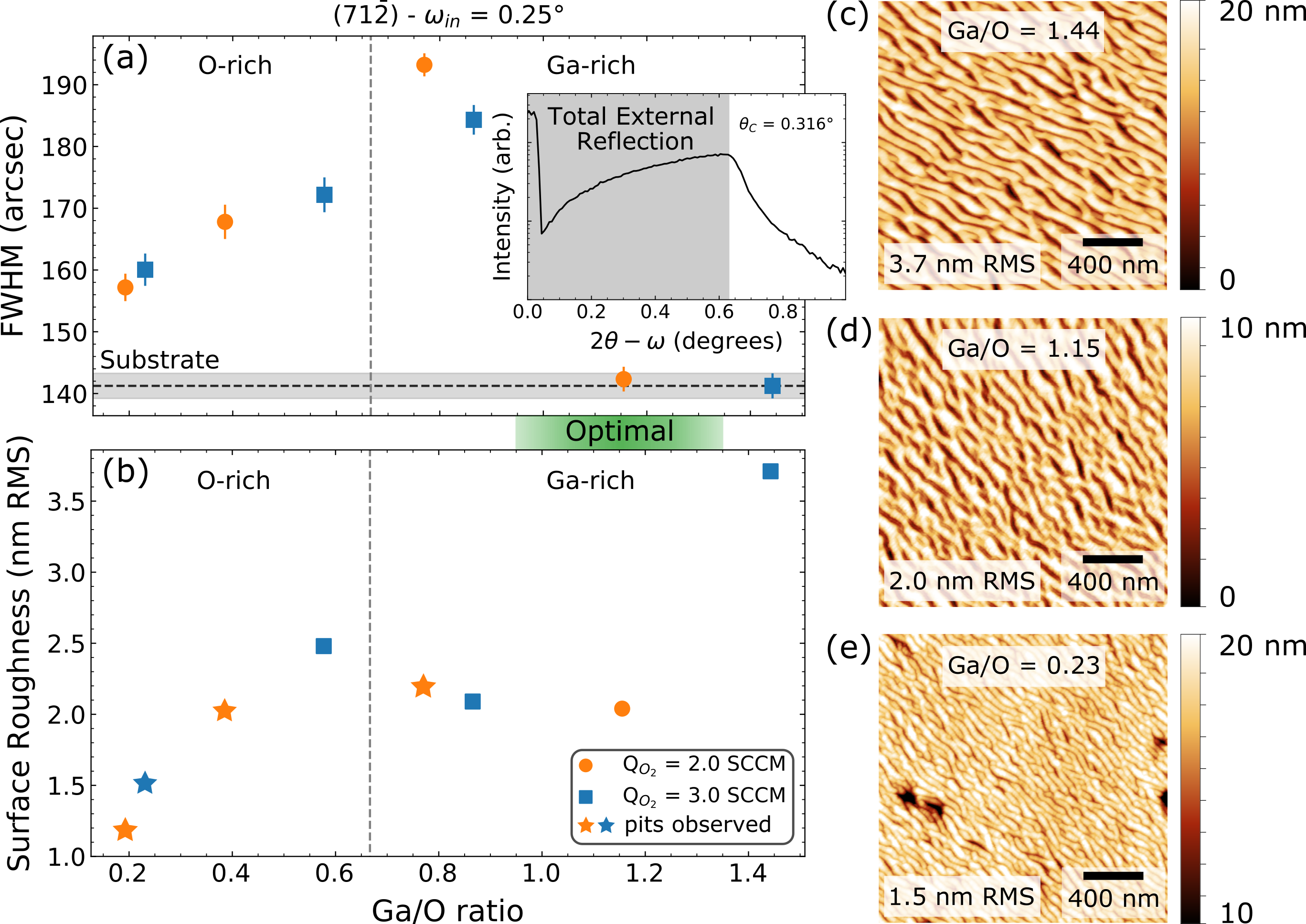}
    \caption{(a) Skew asymmetric XRD rocking curves and (b) surface roughness of $\beta$-Ga$_2$O$_3$ epilayers. (a) $\beta$-Ga$_2$O$_3$ epilayers grown at various flux ratios analyzed by skew asymmetric XRD at an incidence angle of 0.25\si{\degree}, below the critical angle for total external reflection. The vertical bars as well as the shaded grey bar indicate fit error. Critical angle determination by X-ray reflectivity (XRR) is shown in the inset of (a). The crystal quality is observed to decrease slightly as the Ga/O ratio approaches stoichiometry and then improves to near substrate quality above Ga/O ratios of 1:1. Surface roughness determined by AFM is shown in (b) as a function of Ga/O ratio. At a substrate temperature of 700 \si{\celsius}, step-bunching is observed for films across the Ga/O region investigated here. The size of the bunched terraces increases as a function of Ga/O ratio, along with an increased surface roughness.  Films where pitting was observed are notated by stars. The highest  quality films considering both crystal structure and surface morphology are obtained using slightly Ga-rich conditions around 1.15 Ga/O ratio, indicated by the green shaded region. (c) - (e) are 2 \si{\micro\meter} x 2 \si{\micro\meter} AFM scans of films grown at 1.45, 1.15, and 0.25 Ga/O ratio, respectively. (e) demonstrates the pitting observed at low Ga/O ratio. Note that the scale bar minimum in (e) is set to 10 nm for greater visibility, and scale bar maximum in (d) is set to 10 nm.}
    \label{fig:AFM_GIXD}
\end{figure*}

It is important to note that using this method to qualitatively compare samples is only valid if scattering is limited to the epilayer, or if the films are precisely the same thickness. Analysis of substrates without an epilayer show a similar trend of broadening at low incidence angle, with a FWHM of 141\si{\arcsec} for the lowest measure incidence angle, saturating at approximately 30\si{\arcsec} for higher incidence angles (Figure~\ref{fig:xrd_diagram}~(b), grey squares). This broadening is likely due to a large incident X-ray projection on the surface--a geometric effect of extremely low angles. If the length of the beam projected on to the surface is larger than the lateral coherence length of the diffracting crystal (which can be shortened by microstructure, wafer curvature, or surface non-idealities at low angles) then peak broadening will occur\cite{pietsch_high-resolution_2004}. Thus, in order to eliminate the role of thickness variance between different samples the incident angle is chosen to be below the critical angle for total external reflection, measured at 0.316\si{\degree} for $\beta$-Ga$_2$O$_3$ by X-ray reflectivity (XRR) shown in the inset of Figure~\ref{fig:AFM_GIXD}~(a). In this case evanescent X-rays only penetrate a few \si{\nano\meter} in to the film ensuring that all scattering occurs within the epilayer. To ensure instrumental broadening is not dominating the measured signal at these extreme angles, a (100) oriented Si wafer was measured at the (111) reflection using a grazing-incidence skew-asymmetric geometry with an incidence angle of 0.25\si{\degree} ($\chi$ = 53\si{\degree}). The FWHM of (111) Si using the same optics and similar geometry was measured at 18\si{\arcsec}, indicating that instrumental broadening is not a factor in these measurements.

\subsection{Structure and morphology}
Figure~\ref{fig:AFM_GIXD} (a) shows the FWHM as a function of Ga/O ratio for for two oxygen flow conditions,  based on skew asymmetric rocking curves performed along the (71$\bar{2}$) reflection fit to a Voigt profile. The Ga/O ratio was determined using the knee in the growth rate versus Ga flux curve as the stoichiometric Ga flux, shown in Figure~\ref{fig:xrd_stoich}. These XRD measurements were performed in evanescent mode at a grazing incidence angle of 0.25\si{\degree}, below the critical angle for total external reflection, to ensure that the diffraction is sensitive to the epilayer\cite{grigoriev_asymmetric_2016}. Based on Figure~\ref{fig:AFM_GIXD} (a), it appears that the epilayer quality is decreased with increasing Ga/O ratio in O-rich growth regime, and then improved at Ga-rich growth conditions. At the skewed geometry ($\chi$ = 56 \si{\degree}) the incidence angle is a weak function of $\omega$. To further rule out the effects of a changing penetration depth across the independent variable range, peak widths were also analyzed by rotating $\phi$ following the method reported by \textit{Grigoriev et. al.}\cite{grigoriev_asymmetric_2016}, which produced the same qualitative result. A bare substrate was also analyzed (shown by the horizontal dashed line in Figure~\ref{fig:AFM_GIXD}~(a)), giving a baseline broadened value of 141\si{\arcsec} at 0.25\si{\degree} incidence angle, indicating that films grown at the highest Ga/O ratios are of equal quality to the substrate.

\begin{figure*}
    \centering
    \includegraphics[width=17cm]{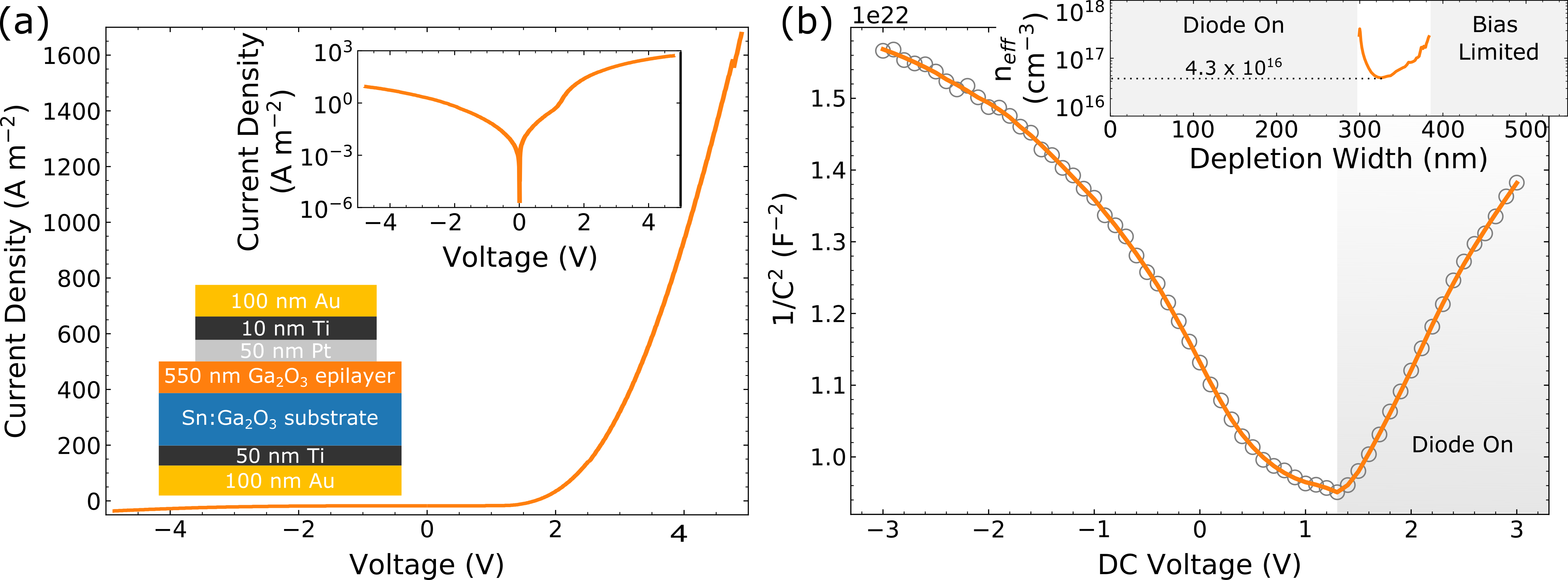}
    \caption{ a) Linear \textit{J-V} curve from a 0.2 mm diameter Schottky diode fabricated on homoepitaxial MBE-grown $\beta$-Ga$_2$O$_3$, grown with a ratio of Ga/O of 0.865. Top Inset: Semilog plot of current density as a function of applied voltage. Bottom inset: the fabricated device stack. Pt is the Schottky contact, while Ti is the Ohmic contact. b) Capacitance-voltage measurement on this Schottky barrier diode plotted as \textit{1/C$^2$-V} for Mott-Schottky analysis. The smoothed curve used to determine the effective carrier concentration n$_{eff}$ is shown by the orange trace. Inset: Calculated effective carrier concentration versus depletion width into the sample, assuming a relative dielectric constant $\epsilon_r$ = 11, indicating an n-type carrier concentration in the homoepitaxial film of \num[scientific-notation = true]{4.3e16} \si{\per\centi\meter\cubed} - \num[scientific-notation = true]{2e17} \si{\per\centi\meter\cubed}. The measurable depletion width is limited by diode turn-on and applied negative bias.}
    \label{fig:Schottky}
\end{figure*}

Surface morphology of the $\beta$-Ga$_2$O$_3$ films was evaluated by AFM as a function of Ga/O ratio during the growth. Surface roughness as a function of Ga/O ratio for films grown at oxygen flows of 2.0 SCCM and 3.0 SCCM is shown in Figure~\ref{fig:AFM_GIXD}~(b) and representative 2 \si{\micro\meter} x 2 \si{\micro\meter} AFM images are shown in Figure~\ref{fig:AFM_GIXD} (c-e).  In agreement with previous reports, the surface morphology for films grown $\geq$ 700 \si{\celsius} consist of step-bunching along terraces which propagate along the $<$100$>$ direction\cite{sasaki_growth_2014}. In Ga-poor growth conditions, adatom diffusion is limited and incorporation occurs before larger terraces can form. For these thin films, all less than 200 nm thick, this results in a lower surface roughness, however that would not hold true for thicker films grown at the same conditions. Low adatom diffusion over the course of a long growth will inevitably lead to roughened growth surfaces--a trend we have observed by RHEED. Films at low Ga/O ratios were also pitted (Figure~\ref{fig:AFM_GIXD} (e),  as annotated by stars in Figure~\ref{fig:AFM_GIXD}~(b). With increasing Ga/O ratio, the scale of the step-bunching increases significantly leading to larger terraces and higher surface roughness (Figure~\ref{fig:AFM_GIXD}~(c)). In the region around stoichiometric growth (2:3 Ga/O ratio) there is very little deviation in the observed surface roughness. For very Ga-rich films the roughness is larger, likely due to partial etching by volatile gallium sub-oxides reported in literature\cite{vogt_competing_2015}. Therefore, the best surfaces are achieved at slightly Ga-rich conditions (Figure~\ref{fig:AFM_GIXD}~(d)) to minimize surface roughness while also minimizing broadening observed by XRD, indicated by the green bar in Figure~\ref{fig:AFM_GIXD}~(a-b).

Based on combined XRD structural quality and the AFM morphology results it is concluded that slightly Ga-rich films, with a Ga/O ratio of approximately 1.15, produce the highest  quality $\beta$-Ga$_2$O$_3$ epilayers for the studied growth temperature of 700 \si{\celsius}. According to XRD alone, the best conditions are Ga-rich (Ga/O $>$ 1.2) and according to AFM the best conditions are Ga-poor (Ga/O $<$ 0.4) followed by slightly Ga-rich (Ga/0 = 0.8 - 1.2); however as noted above this roughness trend may not be true for thicker films. The conclusion about the optimal Ga/O $\approx$ 1.15 growth conditions would not be possible to achieve by either of these techniques alone, highlighting the importance of comprehensive characterization of homoepitaxial layers.

\subsection{Electrical characterization}
To further study the quality of the homoepitaxially grown $\beta$-Ga$_2$O$_3$, vertical Schottky barrier diodes based on thicker epilayers (550 nm) grown with a Ga/O ratio of 0.3 were fabricated and electrically characterized. As shown in Figure~\ref{fig:Schottky}~(a), the fabricated diodes showed current rectification, confirming the formation of Schottky and Ohmic contacts. Fitting to the linear portion of the \textit{J-V} curve, the turn-on voltage is around $V_{TO}$ = 1.4 \si{\volt}, and the on-state resistance is around 230 $\Omega-cm^2$. The shape of the \textit{J-V} curve is non-ideal, and the on/off ratio at $\pm$ 2 \si{\volt} is quite low, around \num[scientific-notation = true]{1e2}. More ideal diode behavior and larger on/off ratios were obtained for thinner samples grown at higher Ga/O ratios, consistent with conclusions from combined XRD and AFM analysis. These results indicate that further optimization of the device fabrication process is needed to improve device characteristics, however these non-ideal Schottky barrier diodes were used for capacitance-voltage measurements due to the thicker epilayers.

Capacitance-voltage data for the  Schottky barrier diode shown in Figure~\ref{fig:Schottky}~(a) was subject to Mott-Schottky analysis, as shown in Figure~\ref{fig:Schottky}~(b).  From this data, a doping density was calculated assuming a relative dielectric constant $\epsilon_{r}$ = 11\cite{Fiedler_2019}, with the results  shown in the inset of Figure~\ref{fig:Schottky}~(b).  The effective n-type carrier concentration in the homoepitaxial layer at 320 \si{\nano\meter} depth from the surface was determined to be approximately  \num[scientific-notation = true]{4.3e16} \si{\per\centi\meter\cubed}, which is significantly lower than for the Sn-doped substrate, N$_D$ - N$_A$ = \num[scientific-notation = true]{1e18} \si{\per\centi\meter\cubed} - \num[scientific-notation = true]{9e18} \si{\per\centi\meter\cubed}. The depletion width was probed from approximately \SIrange{300}{380}{\nano\meter}, limited on the low side by diode turn-on and on the high side by the applied negative bias. The up-turn in the measured carrier concentration below 320 \si{\nano\meter} is due to the positive (but below turn-on) applied DC bias accumulating carriers.  These results further confirm the high quality of the homoepitaxial $\beta$-Ga$_2$O$_3$ layers reported in this study.

\section{
Summary and Conclusions}

In conclusion, we have grown and characterized homoepitaxial $\beta$-Ga$_2$O$_3$ thin films, focusing on the structural and morphological characteristics of epilayers grown at different Ga/O flux ratios. All defect concentrations are below the SIMS sensitivity limit, with exception of In accumulation at substrate-epilayer interface which is due to the sample mounting scheme. Skew asymmetric XRD at grazing incidence was investigated as a method to characterize the structural quality of homoepitaxial layers, demonstrating the effect of incidence angle on measured film quality as determined by rocking curve analysis. Using grazing incidence angles to selectively analyze epilayer structural quality by operating below the critical angle for total external reflection, we have shown that film quality is improved at Ga-rich growth conditions to the point where the epilayer is of equal structural quality to the substrate. Complementary AFM characterization of the surface reveals increased  roughness and step-bunching at increased Ga/O ratios, with films ranging from \SIrange[tophrase={ -- }]{1.2}{3.7}{\nano\meter} RMS. Informed by the combination of AFM and skew asymmetric XRD results, it is determined that optimal growth conditions occur at Ga/O ratios of approximately 1.15, when growing at 700 \si{\celsius} using MBE, however further optimization is needed fine-tune the optimal ratio. Finally, fabricated Schottky barrier diodes characterized by \textit{C-V} measurements reveal an unintentional doping density of approximately \num[scientific-notation = true]{4.3e16} \si{\per\centi\meter\cubed} - \num[scientific-notation = true]{2e17} \si{\per\centi\meter\cubed} in the epilayer, consistent with SIMS depth profiles showing no impurities above the noise level. These results show that it is important to use several complementary measurement techniques to characterize the quality of the $\beta$-Ga$_2$O$_3$ homoepitaxial layers for power electronic applications.

\ack
This work was authored by the National Renewable Energy Laboratory (NREL), operated by Alliance for Sustainable Energy, LLC, for the U.S. Department of Energy (DOE) under Contract No. DE-AC36-08GO28308. Funding provided by the Laboratory Directed Research and Development (LDRD) Program at NREL (growth and measurements), and by the Office of Energy Efficiency and Renewable Energy (EERE), Advanced Manufacturing Office (device fabrication and characterization). The authors would like to thank Bobby To for obtaining AFM images. The views expressed in the article do not necessarily represent the views of the DOE or the U.S. Government.

\section*{References}
\bibliography{report} 
\bibliographystyle{iopart-num} 

\end{document}